\documentclass[12pt,preprint]{aastex}
\usepackage{booktabs}
\usepackage{natbib}

\def\hip{$Hipparcos$}
\def\akari{$AKARI$}
\def\fis{$AKARI/FIS$}

\def\iras{$IRAS$}
\def\iso{$ISO$}
\def\spiz{$Spitzer$}
\def\mass{$2MASS$}
\def\wise{$WISE$}

\begin{document}

\title{Bright Debris Disk Candidates Detected with AKARI/Far-Infrared Surveyor (FIS)}
\author{Qiong Liu\altaffilmark{1}, Tinggui Wang\altaffilmark{1}, Peng Jiang\altaffilmark{1}}
\altaffiltext{1}{Key Laboratory for Research in Galaxies and Cosmology, University of
Science and Technology of China, Chinese Academy of Sciences, Hefei, Anhui 230026, China;
jonecy@mail.ustc.edu.cn; twang@ustc.edu.cn}

\begin{abstract}

We cross-correlate \hip\ main-sequence star catalog with \fis\ catalog, and identify
136 stars (at $>90$\% reliability) with far-infrared detections at least in one band.
After rejecting 57 stars classified as young stellar objects, Be stars and other type stars
with known dust disks or with potential contaminations, and 4 stars without infrared excess
emission, we obtain a sample of 75 candidate stars with debris disks. Stars in our sample
cover spectral types from B to K with most being early types. This represents a unique sample
of luminous debris disks that derived uniformly from an all sky survey with a spatial resolution
a factor of two better than the previous such survey by \iras. Moreover, by collecting the
infrared photometric data from other public archives, almost three quarters of them have
infrared excesses in more than one band, allowing the estimate of the dust temperatures.
We fit the blackbody model to the broad band spectral energy distribution of these stars to
derive the statistical distribution of the disk parameters. Four stars with excesses in four
or more bands require a double blackbody model, three of them are clustered around (100, 40)K
and the other ($\thicksim$200, 50)K.

\end{abstract}


\section{Introduction}

Our solar system is a debris system with the asteroid belt at 2 - 3.5 AU and
the Kuiper belt at 30 - 48 AU \citep{kim05}. Debris disks have been detected in
extra-solar stellar systems as well, commonly referred to as ``The Vega Phenomenon"
\citep{sil00}. The stars with debris disks are generally much older than 10 Myr
\citep{kri10}, which is much longer than the typical time scale of collisional
destruction of dust grains or of spiraling inward due to Poynting-Robertson
drag. Thus dust grains have to be continuously replenished by collisions and/or
evaporation of planetesimals \citep{bac93,wya08}. The studies of debris systems are
significant because they provide a better understanding of the formation and
evolution of planetesimal belts and planetary systems \citep{zuc04,moo11,ray11,ray12}.

The first extra-solar debris disk was detected by the \textit{Infrared Astronomical
Satellite} (\iras) around Vega in 1983 \citep{aum84}. Up to now, nearly a thousand
debris disks have been detected. Most of these systems were found through the
detection of infrared (IR) excess over the stellar photospheric emission. The IR excess
is explained as the dust re-radiation of the absorbed starlight. At the sensitivity
level of the Multi-band Imaging Photometer (MIPS) on \spiz\ \citep{wer04,rie04}, the
incidence of debris disks around main-sequence (MS) stars is about 15\% \citep{kri10}.
Due to their small sizes, only dozens of debris disks around nearby stars have follow-up
direct imaging observations in optical, mid-infrared (MIR), and sub-millimeter bands
(e.g., Schneider et al. 2001; Greaves et al. 2005; Wyatt et al. 2005; Kalas et al. 2006;
Su et al. 2008; Lagrange et al. 2012; also see \footnote{\it http://www.circumstellardisks.org}).

The observed debris disks display diverse properties. Most debris disks have relatively
low dust temperatures between 30-120 K, corresponding to disk sizes from several tens
to a hundred AU for type A to K stars \citep{chen06,moo11,pla09,rhe07}. A small subset of
warm debris disks have been discovered recently with \akari, \spiz\ and
\textit{Wide-field Infrared Survey Explorer} (\wise) \citep{fuj10a,mey08,fuj10b,fuj13,
olo12,rib12}, and the incidence of such disks drops very rapidly with the age of the
stars \citep{urb12}. More recently, \textit{Herschel} revealed a population of cold
debris disks extending to more than one hundred AU with its good sensitivity to the long
IR wavelength \citep{eir11}. A single blackbody model usually provides a
good fit to the MIR spectrum, suggesting that grains are distributed over a relatively
narrow annulus \citep{sch05,chen06}. The relatively narrow width has been confirmed
for some debris disks by direct imaging in the IR and sub-mm bands \citep{boo13}.

The incidence of debris disks as a function of other stellar parameters is of great
interest as it gives further clue to its origin. It appears that the frequency of stars
with debris disks is larger among earlier type of stars, and decreases with the increases
of stellar age \citep{rhe07,wya08}. The rate appears to correlate with the presence of
planets, but not the metallicity of the host stars \citep{mal12}. The general trend
with stellar age reflects the consumption of planetesimals during the system evolution.
However, interpretation of the correlation with stellar types may be more complex since early
type stars have much shorter life-time than late type stars and the correlation may
be entirely caused by the age-dependence of incidence. In addition, the detected
debris disks displayed a wide range of IR excesses from $10^{-6}$ up to $10^{-2}$ of
stellar bolometric luminosity. Over such a wide range, different mechanisms of debris disk
may operate, thus it would be interesting to examine the incidence at a certain fraction of
IR excesses. To explore a large parameter space, a large unbiased sample of debris
disks with known host parameters is required.

Up to now, debris disks have been discovered mostly based on the IR data from four
satellites: \iras\ \citep{man98,rhe07}, \textit{Infrared Space Observatory} (\iso)
\citep{kes96,oud92,abr99,hab99,faj99,spa01,dec03}, \spiz\ \citep{bei06,bry06,chen05,kim05,
moo06,moo11,reb08,rie05,sie07,su06,wu12} and \textit{Herschel} \citep{mat10,eir13}.
\iras\ contained a cryogenically cooled telescope orbiting above the Earth's atmosphere
to make an unbiased all-sky survey at 12, 25, 60, and 100 \micron\ \citep{neu84} at a
relatively poor spatial resolution ($4\arcmin - 5\arcmin$) and
sensitivity (0.6 Jy at 60 \micron\ in Point Source Catalog, 0.225 Jy at 60\micron\ in
Faint Source Catalog)
\footnote{See IRAS Explanatory Supplement, Assendorp et al. 1995, Allam et al. 1996}.
\spiz\ and \iso\ possess much better spatial resolutions and sensitivities than \iras\ but cover
much smaller area of sky at mid and far-infrared (FIR) bands. Thus the latter missions
discovered more faint debris disks. The latter satellites also carried pointed
observations of nearby bright stars that were sensitive to the IR excess down to $10^{-6}$ of
host star luminosity.

In this paper, we search systematically for debris systems around MS stars
by cross correlation of the \hip\ catalog \citep{per97} with AKARI/Far-Infrared Surveyor
\citep{kaw07} All-Sky Survey Bright Source Catalogue (AKARIBSC, Yamamura et al. 2010).
\fis\ surveyed all sky at FIR with a spatial resolution (48 \arcsec) better than
\iras\ and at a sensitivity (0.55 Jy in 90 \micron) comparable to \iras.
The higher resolution will significantly reduce the false contamination in comparison with \iras.
The same with the \iras\ studies, our work also focuses on the IR bright debris disks that
complement the deep surveys from \iso\ and \spiz.
Our primary motivation is to search for FIR excess stars by \fis\ and discuss the fundamental
parameters of the disks such as dust temperature, fractional luminosity and dust location.
These parameters can be estimated from the spectral energy distribution (SED) of dust emission.
Fortunately, all IR excess stars except HIP 57757 in our sample have \wise\ detections which
lead to better wavelength coverage than many previous searches. As shown by Moor et al. (2011), the
interpretation of a SED is ambiguous, but by handling a debris disk sample as an
ensemble, one can obtain a meaningful picture about the basic characteristics of the
parent planetesimal belt(s) and about the evolutionary trends. The paper is arranged as follows.
We will describe the data sets and methods used in the
construction of the debris disk sample in Sect. 2; and present an analysis of the properties
of the disks as well as their host stars in Sect. 3; In Sect. 4, we discuss the sample
comparison; Finally, we present the conclusion in Sect. 5.

\section{The Method and the Sample}

\subsection{Matches between \hip\ catalog and AKARIBSC}

\begin{figure}
\plotone{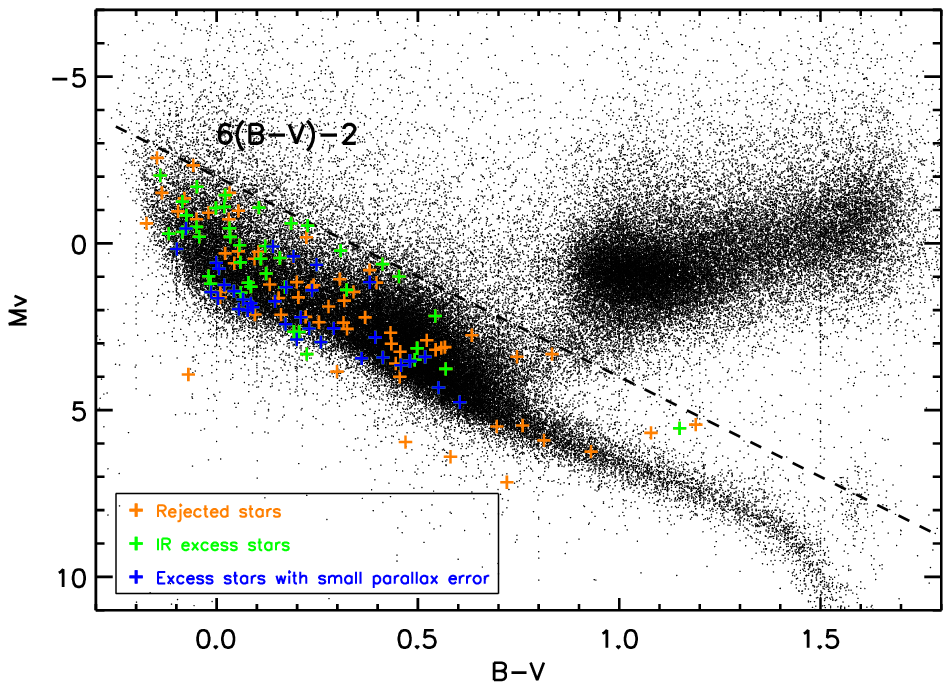}
\caption{Selection of main sequence (MS) stars on H-R diagram of the \hip\ field stars.
The stars below the dashed line are MS stars, which have been searched for far-infrared
excess emission using \fis. The FIR excessive stars are plotted with a plus: green
and blue plus represent for the debris disk candidates, blue plus for the source with an
accuracy in the parallax measurement to 10\%, orange plus for the rejected source.
}
\label{mvbv}
\end{figure}

The primary star catalog used in this work is the \hip\ catalog, which contains over 110,000
stars with precise photometry as well as astrometry of unprecedented accuracy for the nearby
stars \citep{bes00}. In Figure \ref{mvbv}, we show a Hertzsprung-Russell (H-R) diagram for all
the cataloged stars by extracting the colors (\bv) and parallaxes from \hip\ database.
The MS stars are selected according to the criterion $M_{V} \geqslant 6.0(\bv)-2.0$
\citep{rhe07}. This results in a catalog of 67,186 \hip\ MS stars.

We then cross-correlate the catalog with the AKARIBSC to identify the \hip\ MS stars
detected in the \fis\ bands. Since the AKARIBSC has much worse position precision than
the \hip\ catalog, we determine the matching radius based on the performance of \akari\
only. The spatial distribution of \fis\ sources is very inhomogeneous on the sky, so a
uniform matching radius is not an ideal choice. To show this, we write the false detection
rate for a sub-sample of stars on the sky with the background surface density $n$ of IR sources
as follows,
\begin{equation}
R_{false}=\frac{N_{false}}{N_{total}}=\frac{N_* \pi r^2 n}{N_*fc(r)+N_*\pi r^2 n}=
\frac{\pi r^2 n}{fc(r)+\pi r^2 n}
\label{falsematching}
\end{equation}
where $f$ is the fraction of \hip\ stars with IR fluxes above the detection limit,
$N_{total}$ and $N_{false}$ are the number of all matches and the expected number of
chance matches. $c(r)$ is the completeness with a matching radius $r$, i.e., the
probability of a real matching source falling within a circle of radius $r$ around
the star, which is determined by the position error ellipse of the IR source. Assuming
$n$ does not correlate with $f$, the fraction of false matching increases with the
background surface density of IR sources at a given matching radius. In reality, $f$
and $n$ might be correlated, e.g., young stars are more likely located on the Galactic
plane, where stellar surface density is also higher; as such the false matching fraction
may not follow Eq. \ref{falsematching} exactly. Anyway, we will determine the matching
radius according to the surface density of IR sources. Since $c(r)$ increases slower
than $r^2$, as $r$ increases, the false matching rate increases.

\begin{figure}
\plotone{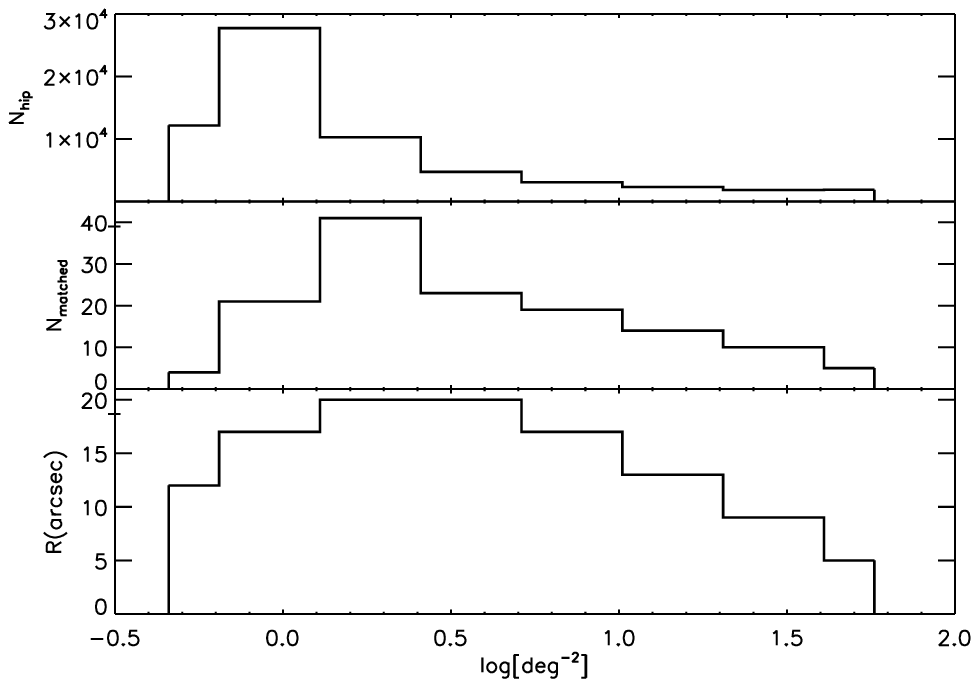}
\caption{The \fis\ local surface density distribution of the \hip\ MS stars and the
corresponding matching radius and matched numbers.}
\label{density}
\end{figure}

As a trade-off between the reliability and completeness, the matching radius at a given
surface density is so chosen that false detection rate is less than 10 \%.
In practice, we estimate the local \fis\ source density around each \hip\ star, and then divide
\hip\ stars into different density bins. For each bin, we increase the matching radius iteratively
from 5 \arcsec, to a radius where the false detection rate is close to 10\%
or to the upper limit of 20 \arcsec. The cross-correlation results in 136 matching pairs. Figure \ref{density} presents the number of \hip\ stars (upper panel), matched IR sources (middle panel), and matching
radius (bottom panel) at each bin of the local IR source density. An interesting feature in this
plot is that the peak of matched pair distribution is shifted to the high density area rather than
to lower density area as expected. This implies a strong correlation between $f$ and $n$.
Regions of lower density have a lower fraction of stars with bright debris disks perhaps because
the chance of finding young stars in such regions is lower. While at higher density regions, the
higher excess fraction in the Galactic plane may be justifiable and the magnitude of this effect
can be estimated from Figure 2 which looks like the fraction is fairly constant above
$\sim1\;deg^{-2}$, but a factor of 6 lower for lower far-IR densities.

Next, we remove the sources with obvious contaminations in the IR. Seven stars in nebula
are rejected because nebula is a FIR source. Among them, three stars are in Kalas's
sample \citep{kal02}; nebula is clearly seen in the images of other three stars returned by SIMBAD;
and one additional star (HIP 78594) was rejected by Mo\'or et al. (2006) based on the image of the
Digitized Sky Survey, which shown a reflection nebulosity around this star. Another contamination source
is the emission from cold diffuse interstellar dust (cirrus, Rhee et al. 2007), which also emit MIR
(e.g., Boulanger et al. 1998). We reject the cirrus contamination stars based on their MIR images obtained by
\wise\ \citep{wri10}. \wise\ has mapped the whole sky in four IR bands $W1$, $W2$, $W3$
and $W4$ centered at 3.4, 4.6, 12 \& 22 \micron\ with 5 $\sigma$ point source sensitivities better
than 0.08, 0.11, 1 and 6 mJy, respectively. The angular resolutions are 6\farcs1, 6\farcs4, 6\farcs5
\& 12\farcs0 at corresponding bands, and the astrometry precision for high SNR sources is
better than 0\farcs15 \citep{wri10}. The high sensitivity and high angular resolution images
are used to remove the confusion source and to further constrain the disk properties in the
SED fitting. All stars except HIP 57757 are covered by \wise. We check the \wise\ images of these
sources for the presence of weak diffuse emissions around stars. Seven stars are affected by potential
cirrus emissions and rejected, leaving 122 stars for further study. Note that
the number of rejected contaminated sources is in agreement with the expected chance
matches.

\subsection{Infrared Emission of Stellar Photosphere}

Obtaining the flux densities of stellar photosphere is essential for identifying and
measuring the strength of an IR excess \citep{bry06}. We collect the optical to
near-infrared (NIR) absolute photometric data of stars in our sample to construct SED.
Optical magnitudes in $B$ and $V$ are taken from the \hip\ satellite measurements.
NIR photometries $JHK_s$ are extracted from Two Micron All Sky Survey (2MASS) catalogs
\citep{skr06}. The observed magnitudes are converted into flux density (Janskys) using
the zero magnitudes in Cox (2000) \citep{rhe07}.

The stellar SEDs are fitted with the latest Kurucz' models (ATLAS9)
\footnote{\it http://wwwuser.oat.ts.astro.it/castelli/grids.html} \citep{cas04}.
The models cover wide ranges of four parameters: temperature, surface gravity,
metallicity, and projected rotational velocity. For each stellar type, we select only
a subset of model spectra from ATLAS9 according to Allen's astrophysical quantities
\citep{cox00}. For B-type stars, the effective temperatures are in 500~K increments
from 10000 to 20,000~K, the surface gravity log~$g$~cm~s$^{-2}$ value is 4.0.
For A-type and later types stars, the effective temperatures are in 250~K increments
from 3,500 to 10,000~K, the surface gravity log~$g$~cm~s$^{-2}$ values are 4.0, 4.5.
We chose microturbulent velocity $\xi$=2~km~s$^{-1}$ and metallicity value
$[M/H] =0$ (solar metallicity) for all cases.

We fit the model spectra to the observed SEDs from optical to NIR for each object in
order to find the best matched stellar models. During the fit, the stellar spectra are
reddened and convolved with the response of each filter to yield the model flux density
at each band. This method gives the model flux density more accurately than adopting
a constant magnitude to flux conversion factor, especially when the passband includes
significant spectral features such as the Balmer jump \citep {rhe07}. For each stellar
model, the best fit is obtained by minimizing $\chi^2$ with the extinction $E(B-V)$ and
normalization as free parameters. We select out the best model with the smallest $\chi^2$
among different stellar models. Using the best-fit Kurucz model, we estimate the stellar
photospheric flux densities in the \wise\ and \akari\ bands.

\begin{figure}
\plotone{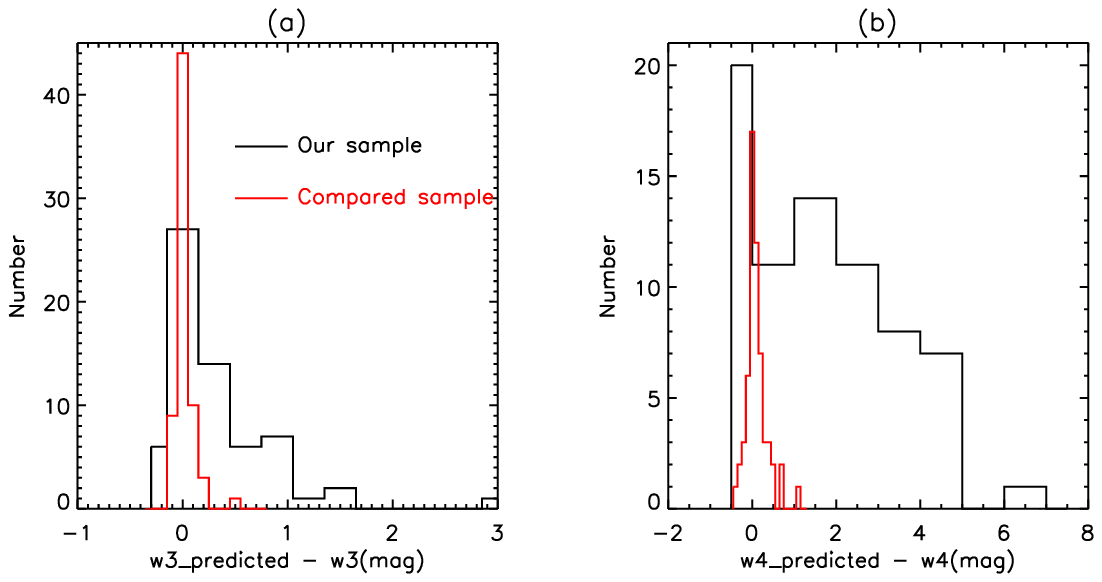}
\caption{The distribution of the difference between the observed magnitude in \wise\
$W3$ and $W4$ bands and the predicted stellar photosphere model. The black line
represents for our final sample. The red line is for the matched random sample
as described \S2.2.}
\label{wisehist}
\end{figure}

To assess the reliability of stellar photospheric flux predicted by the best model in
the \wise\ $W3$ and $W4$ bands, we examine the distribution of the differences between
observed and predicted magnitudes for a sample of randomly selected \hip MS stars,
which usually should not show MIR excesses. The sample is so compiled that
the comparison sample well matches our final debris disk sample in the distribution
of Galactic latitudes, stellar spectral types as well as their optical magnitudes.
The size of the comparison sample is a factor of two larger. We fit the optical to
NIR photometric data of the comparison sample with stellar models as described above.
The distributions of the differences are fairly narrow with almost no systematical
offsets (Figure \ref{wisehist}): $<W3(observed)-W3(model)>\simeq 0.002$mag,
$<W4(observed)-W4(model)>\simeq 0.04 $ mag, and $\sigma(W3)=0.06$ mag and
$\sigma(W4)=0.13$ mag. In the following analysis, we will incorporate these numbers
as the systematical uncertainties of model fluxes in the two \wise\ bands.

\subsection {Identification of Debris Disk Candidates}

Our goal is searching for the IR excess from debris disks, while debris disk is not
the only source of the IR excess. So we will remove other IR excess sources from our
sample. Firstly, in some young O stars, significant IR excess may arise from gas
free-free emission instead of from the debris disk. These stars generate strong ionized
winds that produce strong IR and radio excesses. Thus five O stars are excluded from our
sample. Secondly, a Be star is a B-type star with prominent hydrogen emission lines in
its spectrum and IR excess \citep{por03}. Both emission lines and excessive IR emission
in Be stars are formed in the circumstellar disks, that are most likely ejected or
stripped from the stars themselves. We rejected 12 Be stars based on SIMBAD classification.
Thirdly, we reject 3 objects including a star without reliable flux density of \fis (none
of the band has the quality flag=3), a quasar and a Post-AGB stars.

Finally, young stellar objects (YSOs) often harbor protoplanetary disks \citep{moo06}, and
also display IR excesses. We will reject them according to the shape of their SEDs in the
IR as follows. YSOs are classified observationally according to the shape of their SEDs in
the IR between the K band (at 2.2 \micron) and the N band (at 10 \micron) defined as (Armitage 2007),
\begin{equation}
\alpha_{\rm IR} = \frac{\Delta \log (\lambda F_\lambda)}{\Delta \log \lambda},
\end{equation}
where $\alpha_{\rm IR}>-1.5$ is a strong indication for a YSO. In our sample, several
stars have the YSOs' SED features as: Class I (approximately flat or rising SED into
mid-IR ($\alpha_{\rm IR} > 0$)) and Class II (falling SED into mid-IR
($-1.5 < \alpha_{\rm IR} < 0$)). Class I YSOs are typically younger and possess more
massive disks than Class II objects. In principle, YSOs should have been removed from
our selection of MS stars using the H-R diagram. However, stars would cross the
MS belt on the H-R diagram when they evolve from pre-main-sequence (PMS) to the
zero-age main sequence (ZAMS) stars. Most of these stars are very close to ZAMS, and only
a small fraction may have massive planetary disks. According to our SED fitting, we reject
20 YSOs in total and list them in the Table \ref{tb-reject}. In addition, we also purge
another 3 stars (HIP 53911, HIP 77542, HIP 23633) classified as YSOs in SIMBAD although
their IR slope does not meet above criterion. All these rejected stars are listed in Table 1.
We retain a sample of 79 stars.

In order to assess whether there is an excess IR emission in the rest of the sample,
we calculate the significance of the excess to the stellar photospheric emission model
in each \fis\ band using following formula \citep{bei06,moo06}:
\begin{equation}
\chi = [F_{IR} - F_{phot}] / \sigma_{IR}
\end{equation}
where $F_{IR}$ is the measured flux density; $F_{phot}$ is the predicted photospheric
flux density; and $\sigma_{IR}$ is the uncertainty of the measured flux density.
An object is considered as an excess candidate star when $\chi > 3.0$ \citep{su06}
in one or more of 65, 90, 140 or 160 \micron\ bands. Applying this criterion, we
identify in total 75 FIR excess stars in the \fis\ database. Due to the shallow
\fis\ flux limit, only 4 of the Hipparcos stars were sufficiently bright
to have their photospheres detected in the far-IR in the absence of a FIR excess.
Among these 75 stars, 72 stars have high quality 90 \micron\ flux densities
($fqual=3$). The other three were flagged as having unreliable 90 \micron\ fluxes.
Two of them are safely detected in 140 \micron\ or 160 \micron\ bands, indicating the
presence of a cold disk; and the third has reliable fluxes in both 65 and 140
\micron\ and is a well known bright debris source.

The MIR excesses from WISE 22 \micron\ and 12 \micron\ are estimated in the same way.
Among the 75 objects, 53 stars show excesses in the 22 \micron\ band at more than 3$\sigma$ level
(see Figure \ref{wisehist}b) after considering the systematical uncertainty of \textbf{0.13}
mag (\S2.2). 37 stars show excesses in the 12 \micron\ band after considering the systematical
uncertainty of \textbf{0.06} mag.
The \wise\ magnitudes for these 75 objects are presented in Table \ref{tb-flux}.

\section{Properties of Debris Disks and Host Stars}

We have identified a sample of 75 stars with debris disks. In this section, we will study
the properties of host stars and debris disks. The stellar properties include magnitude,
color, location on the H-R diagram as well as those derived from the SED fitting in
the previous section. The properties of debris disks are derived by using the parameters
obtained in modeling the IR excesses in \fis\ and \wise\ data.

Previous studies suggested that debris disks are optically thin and usually consist
of a narrow ring \citep{bac93} in thermal equilibrium with the stellar radiation
field. Therefore the IR excess is usually modeled as a single temperature blackbody
\citep{kim05, bry06, rhe07}.
There are two free parameters in the fit, blackbody temperature and its
normalization. To fully determine the model parameters, excesses in at least two bands
are needed, while with more data points, we can get a best fit by minimizing $\chi^2$.
Therefore, according to the number of bands with detected FIR and MIR excesses
(\fis\ 4 bands and \wise\ 12 \micron\ and 22 \micron),  we further divide the IR excess
sample into two groups: IR excess in a single band (Group I) and excesses in two
or more bands (Group II). Note both Group I and II should show excess in at least
one \fis\ band. Only for sources in Group II can the dust temperature be fully
determined for the single temperature dust model, while in Group I, by combining \fis\ data
with the upper limits at the WISE 22 \micron, we can derive an upper limit on the dust
temperature. Among 75 debris disk candidates, the majority (55)\footnote{53 stars with 22
\micron\ excess and 2 stars without 22 \micron\ excess but with two or more \fis\ band
excesses.} are in the Group II. In passing, we note that 11 objects are detected in two or
more \fis\ bands. They are brighter at 90 \micron\ on average, and a significant
fraction (9/11) of these sources does display MIR excess. Similarly, bright sources
are more likely to show 22 \micron\ excesses.

\subsection{Stellar Properties of Debris Disk Hosts}

It is evident that the debris
stars do not evenly sample its parent Hipparcos stars, but are biased to early type stars,
consistent with previous study \citep{rhe07}. It is puzzling that these stars
are not particularly close to the lines of ZAMS, while previous studies suggested that incidence
of debris disk decreases with the increase of the stellar age. The discrepancy
may be caused by three factors, the contamination of PMS stars, large errors in the parallax
measurements and large interstellar reddening. Since PMS stars have been excluded from the sample,
so only the latter two possibilities need to consider. If we only include these sources with
accuracy in the parallax measurement to 10\%, most sources tend to distribute near the
ZAMS (blue plus in Figure \ref{mvbv}), suggesting that the discrepancy is attributed at
least in part to the inaccuracy of the distance estimate. However, the rule of dust extinction
cannot be ruled out because both the extinction and uncertainty of parallax increase with the
distance of stars thus are likely correlated.

\subsection{Disk Properties}

By fitting the IR excess flux densities, we will derive the dust temperature and the fraction
of the stellar luminosity reprocessed by dust. By combining with additional stellar
parameters, we can estimate the dust location and other quantities. The inferred basic disk
properties are listed in Table \ref{tb-dust}. In the following subsections, we will describe
the method and results in detail.

\subsubsection{Dust Temperature}

We fit the excessive flux densities in the \fis\ and \wise\ bands with a single temperature
blackbody model as described, convolved with the response functions of the corresponding
filters. In the case of single band excess (Group I, 20 star in total), we derive a
maximum temperature by combining the excess \fis\ flux with the upper limits at 22
\micron, and the normalization of blackbody radiation at the maximum temperature. Note
that this normalization is usually higher than the one assuming that the blackbody peaks
at the detected IR band, as has been made in Rhee et al. (2007). In the case of two band
excesses, we can fit the blackbody solution directly to determine the temperature and
normalization. We estimate the uncertainty of a parameter by using $\chi^2$ as a function
of the parameter. We adopt $\Delta\chi^2=2.7$ in the error estimate, i.e. at 90\%
confidence level for one interesting parameter. In the case of more than two band excesses
(38 objects including 37 stars with 12 \micron\ excess; 30 stars have three band excesses
and 8 stars have four or more band excesses),
the best fit parameters are determined by minimizing $\chi^2$, and again the uncertainties
of parameters are given at $\Delta\chi^2=2.7$. The typical uncertainty in the dust temperature
is about 3 K. We do not use \iras\ fluxes because these fluxes may suffer from contaminations,
in particular for the objects beyond 100 parsecs, where the contamination of cirrus is
severe due to poor spatial resolution of \iras.

\begin{figure}
\plotone{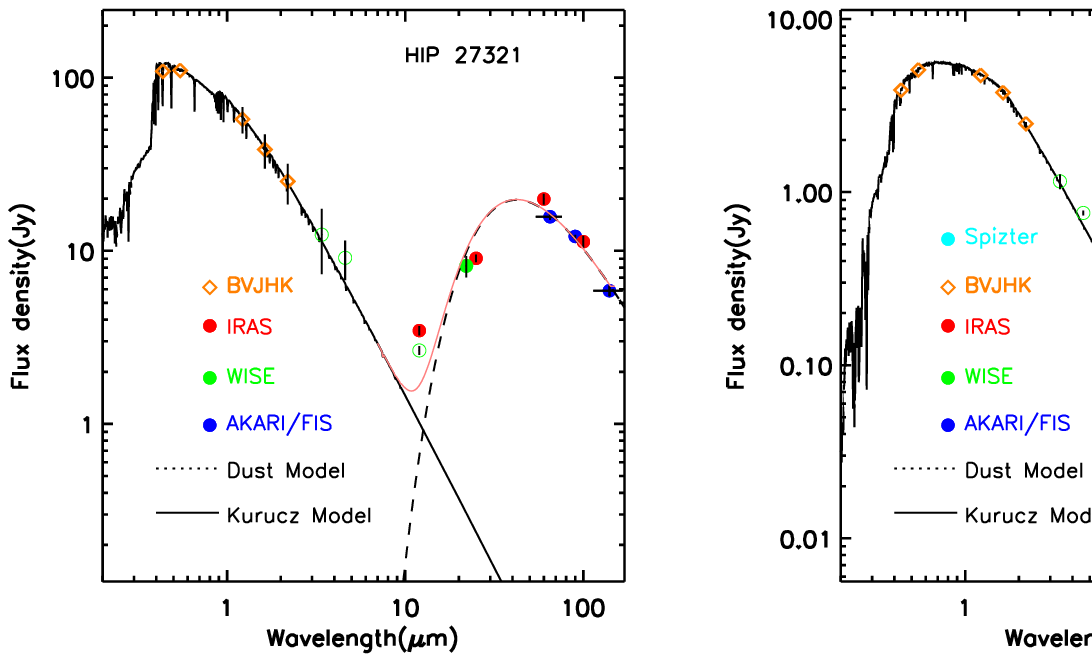}
\caption{SEDs for IR excess stars in our sample. The photospheric models and the disk
models are shown as solid black lines and dotted lines, respectively.
The different symbols represent the different data-sets - orange diamonds: $BVJHK$,
red filled dots: \iras, blue filled dots: \fis, cyan filled dots: \spiz, green filled dots:
\wise\ without saturations, green hollow circles: \wise\ with saturations.}
\label{bbmatch}
\end{figure}

In most cases, a single temperature blackbody usually gives an acceptable fit to the
data for sources with multi-band excesses. Examples of SED fitting are
shown in Figure \ref{bbmatch}. We consider minimum $\chi^2 > 6.7$ for three band excesses
and $\chi^2 > 9.2$ for four band excesses to be unacceptable (at $< 1\%$ probability).
Using these criteria, 15 stars require a more complicated model including
11 stars with excesses in three bands and 4 stars with excesses in four or
more bands. This indicates either multi-temperature components or
a non-blackbody nature of the dust grains. We use a double blackbody model
to fit the SED of the four stars with excesses in four or more bands (Figure
5). The temperatures for the two components are (206, 54) K, (110, 41) K, (107, 45) K
and (105, 37) K, respectively. Three of them are clustered around
(100, 40) K, and the other ($\thicksim$200, 50) K. The star with a warm component ($\thicksim$200 K)
has a fairly large ratio of excesses in $W3$ and $W4$, so the presence of warm
component is independent of modeling, while other three show rather steep
spectrum between $W3$ and $W4$. The temperature of the warm component is
similar to that of grains in the asteroid belt of our solar system, and while
the cold component in the Kuiper belt. For 14 stars with three band excesses
that cannot be well fitted by a single blackbody model, it is insufficient to
constrain the temperatures in a dual blackbody model. However, according to
the excess flux ratio between W3 and W4, we speculate that about 7 stars may
have a warm component.

Note weak 12 \micron\ excess does not significantly affect the fit to the cold component.
Therefore, we will focus only on the cold component in the rest of this paper.
The best fitted $T_{\rm d}$ is listed in column (6) in Table \ref{tb-dust}. The distribution
of $T_{\rm d}$ is shown in Figure \ref{dusthist} (b). Dust temperatures are falling in the
range of 27 to 194 K with a median value of 78 K. The dust temperatures of the disk
correspond to the peak of blackbody emission from 26 \micron\ to 189 \micron.

\subsubsection{Fractional Luminosity}

Fractional luminosity $f_{\rm d}$ is defined as the ratio of IR luminosity of the debris
disk to that of the star, frequently used to characterize the effective optical depth
of the disk,
\begin{equation}
f_{\rm d} = L_{\rm ir}/ L_\star
\end{equation}
where $L_{\rm ir}$ is the IR luminosity estimated by the fitted IR blackbody model.
The stellar luminosity $L_\star$ is calculated from the best-fit Kurucz model.
The uncertainties of $f_{\rm d}$ can be estimated by a combination of the uncertainties
in the temperatures $\sigma_{T_d}$ and normalization. The typical uncertainty of
$f_{\rm d}$ is 0.14 for our sample.

We plot the distribution of the $f_{\rm d}$ in Figure \ref{dusthist}(c). Our sample spends
a large range of $f_{\rm d}$, $1.04\times 10^{-5}<f_{\rm d}<0.06$, with a median
value of $1.18\times 10^{-3}$. Limited by the sensitivity of the \fis, the distribution itself
should not be taken too seriously because disks with low $f_{\rm d}$ can be
detected by \fis\ only for very bright nearby stars, resulting a distribution strongly
biased to the higher $f_{\rm d}$.

\subsubsection{Dust location, Dust mass}

With the assumption that the debris disk is optically thin in thermal equilibrium
with the stellar radiation field, the temperature of a dust grain with a given
chemical composition and grain size depends on the radial distance to the central
star only \citep{kim05}. Assuming that the dust is located in a narrow ring between
$R - {\rm d} R$ and $R+ {\rm d} R$, one can write the radius of dust ring $R$ by the
following formula \citep{bac93}:
\begin{equation}
R_{\rm d} =  (278 / T_{\rm d}) ^{2} (L_\star / L_\odot) ^{0.5}
\end{equation}
Because this formula assumes that the dust is blackbody-like, the resulting $R_{\rm d}$
corresponds to a minimum possible radius \citep{moo11}. The uncertainties of $R_{\rm d}$
are estimated from the error propagation of the uncertainties of temperatures $\sigma_{T_d}$.
This gives typical error of 10\% in $R_{\rm d}$ formally. Figure \ref{dusthist} (d) shows
the distribution of the dust location $R_{\rm d}$.

The total mass \textit{M} of dust can be written by the following formula \citep{rhe07}:
\begin{equation}
M_{\rm d} = (4 \pi a^{3}/3) \rho N
\end{equation}
where \textit{N} is the total number of grains in the disk, $a$ and $\rho$ are the mass
weighted average radius and density of grains. For an optically thin dusty ring/shell of
characteristic radius \textit{R},
\begin{equation}
 f_{\rm d} = N\pi a^{2}/(4\pi R_{\rm d}^{2}).
\end{equation}
Then,
\begin{equation}
 f_{\rm d}/M_{\rm d} \varpropto 1/(\rho aR_{\rm d}^{2}).
\end{equation}
If the characteristic grain size and density do not vary much in different debris disks,
then one expects $f_{\rm d}$/$M_{\rm d}$ to vary as the inverse square power of the disk
radius, $R_{\rm d}$ \citep{rhe07}. The slope is a constant, which can be
derived from the disks that their masses were derived from sub-millimeter data. We use
the slope from Rhee et al. (2007) (see Figure 5). So we can change the equation
to the following form:
\begin{equation}
M_{\rm d}=f_{\rm d} (R_{\rm d}^{2}/9.12) M_\earth.
\end{equation}
where, $f_{\rm d}$ and $R_{\rm d}$ are taken from Table \ref{tb-dust} in column (9) and
column (7) respectively. Then the calculated dust mass is listed in column (8) in Table
\ref{tb-dust}. Noting that we have adopted several assumptions in deriving dust mass, that
are only valid statistically, thus the dust mass is only a rough estimate for individual
object.

\section{Discussion}

An effective way of characterizing the sample is to make a comparison with other
sample in literature. A similar one is the \iras\ debris disk sample, which was
constructed by cross correlating the \hip\ MS star catalog with the
\iras\ Point Source Catalog (PSC) and Faint Source Catalog (FSC) \citep{rhe07}.
The sample consists of 146 stars within 120 pc of the Earth that show excess emissions
at 60 \micron. The distance limit is so set to avoid possible heavy contamination arising
from interstellar cirrus or star-forming regions. Most of these stars belong to early
types, from late B to early K-type stars, similar to our sample. Despite the similar
sensitivity of \iras\ at 60 \micron\ band and \fis\ at 90 \micron, only 27 stars in
Rhee et al. are in common with ours, while 35 stars in total are within
120 pc in our sample.

To understand what causes the difference, we search the flux density at 60 \micron\ of
Rhee's sample from the \iras\ PSC and FSC with a matching radius of 45\arcsec\ as
described in Rhee et al. (2007) Figure \ref{f90} shows that \iras\ flux at 60\micron\
and \akari\ flux at 90\micron\ are fairly well correlated for these detected in both
bands (29 in total including 27 in our sample and 2 PMS stars rejected from our sample
mentioned in \S2.3) and within 120 parsecs of the Earth. A logarithm linear fit to the
data yields a best fit $\log f_{90 \micron}= 0.90 \log f_{60 \micron}-0.08$.
The ratio of the two fluxes is certainly dependent on the disk temperature, but with only
29 data points in hand,
we will not explore this further. In our sample, 27 stars beyond 120 parsecs have \iras\
detection. For these stars, the flux ratios between \iras\ 60 \micron\ and that of
\fis\ 90 \micron\ are substantially higher than those of nearby stars, especially for
sources with $f_{60 \micron}>$ 1 Jy. This is likely caused by source contamination in \iras.
With a factor of more than four improvement in the spatial resolution, the contamination
is greatly reduced in \fis\ flux.

We examine these sources that appear only in one sample in detail. 9 stars within 120
parsecs are included in our sample but not in Rhee et al. (2007).
5 of them (HIP 10064, HIP 20884, HIP 26062, HIP 48613, HIP 57757) have \iras\ detections.
HIP 57757 was rejected by Rhee as a non-IR excess star.
4 stars are not in the rejected source table of Rhee et al.
HIP 20884 and HIP 26062 were excluded due to the spectral type cut in Rhee et al.
HIP 10064 and HIP 48613 were reported to show a FIR excess (Oudmaijer 1992; Chen et al. 2006).
While the rest 4 of them (HIP 65875, HIP 77441, HIP 88983, HIP 90491) do not have available
\iras\ data. The \fis\ 90 \micron\ flux densities of these 4 stars are in
the range from 0.4 to 0.7Jy. Adopting above relation between 90 and 60 $\micron$ relation for
sources in common, 60 $\micron$ flux is expected to be 0.4-0.8 Jy. HIP 88983 and HIP 90491 are
in ``IRAS Faint Source Catalog Rejects" with a 60 $\micron$ flux of 0.33 Jy and 0.28 Jy with
$fqual=2$ and 1, respectively. This suggests that these sources have fainter 60 $\micron$ flux
than expected, which indicates cool debris disks. To produce a flux ratio of 0.6 between
60 to 90 $\micron$, the blackbody temperature would be about 50K.
At this temperature, flux ratios of 140 and 160 to 90 $\micron$ are about 0.9 and 0.7,
so the expected 140 and 160 $\micron$ flux is below the detection limits in both bands,
consistent with none detections in either band. The other two stars may have similar situations.
Note two of debris disk candidates in the sample that have a firm flux only at 140 or
160 $\micron$ are likely even cooler.

On the other hands, most sources (119/146) in Rhee's sample are not in ours.
Most of these sources (113/119) have a flux $f_{60 \micron}<0.63$Jy, and it is not surprise
that they are undetected by \fis (red dots in Figure \ref{f90}). According to the
relation between $f_{90}$ and $f_{60}$, we expect $f_{90 \micron}<0.55$Jy, which is below
the formal flux limit of the AKARI. Thus undetected of these sources may be entirely due to
shallowness of the \fis\ survey. 6 bright \iras\ stars are not in our final sample. Among them,
the two brightest stars (HIP 53911 and HIP 77542) are actually detected by \fis\, but rejected
as PMS according to SIMBAD classification. We checked the other four stars and found no matching
sources in \fis\ catalog even at a larger matching radius.

For comparison, we over-plot the disk parameters of IRAS sample (Table 2 of Rhee et al.
2007) in Figure \ref{dusthist}(b),(c),(d). Only sources with a fitted temperature, i.e.,
detected in more than one \iras\ band, are shown. Our sample distributes in a relatively
narrower temperature range than the \iras\ sample. However, as we have discussed above
that 4 stars in our sample without available \iras\ data may have very low temperature,
then this difference of temperature distribution between two sample may be not real.
Our sample tends to have larger excesses and to possess more distant disks due to the
different flux limits and the distance cut used in selection of the sample.

A caution should be given here that both dust temperature and normalization in Rhee et
al. (2007) is based on solely \iras, which have much poorer sensitivity in the MIR
in comparison with \wise. Therefore, the dust temperature for a large fraction of
objects in their sample could not be determined, and was artificially assigned to
85 K so the peak emission is at 60 \micron. Even for those objects with multi-bands
\iras\ detections, the dust temperatures were less well determined than in this paper.

Finally, a total of 43 stars were already reported in literature (notes in Table
\ref{tb-dust}), so 32 stars are reported to have infrared excesses for the first time
in this paper. Most of them locates at a distance more than 120 parsecs from the Earth,
but are relatively very luminous. As Kalas et al. (2002) pointed out: ``Pleiades-like
dust detected around the star is capable of producing the FIR emission rather than the
Vega phenomenon". HIP\,78594 (Table 1, marked with `a') which was rejected by Mo\'or et
al. (2006), is such a kind of star. So these 32 new IR excess stars need to be further
checked out by coronagraphic optical observations to confirm whether debris disks are
response for the infrared excesses.

\section{Summary}

In this paper we cross-correlate the AKARIBSC with the \hip\ MS star catalog
using a matching radius adapted to the local stellar surface density and yield a sample
136 far-infrared detected stars (at $>90$\% reliability) at least in one band. After
rejecting \textbf{57} stars classified as young stellar objects, PMS stars and
other type stars with known dust disks or with potential contaminations, we obtain a
sample of \textbf{75} candidate stars with debris disks and 4 stars without FIR excess.
The stars in the sample spans from B to K-types, with only 2 G-type and 1 K-type stars.

With the shallow limit of \fis, the survey can only recover the brightest debris disks.
This represents a unique sample of luminous debris disks that are derived uniformly
from an all sky survey with a spatial resolution a factor of two better than the
previous survey by \iras. This sample is also complementary to the deep, small area
surveys or deep surveys of nearby stars as already carried out with \spiz\ and \iso
that find out mostly faint debris systems. Moreover, by collecting the IR photometric
data from other public archives, 55 stars have IR excesses in more than
one band, allowing an estimate of the dust temperature. We fit a blackbody model to
the broad band SEDs of these stars to derive the statistical distribution of the disk
parameters. 4 objects with four or more band excesses can be fitted by a double blackbody
model. Three of them are clustered around (100, 40) K, and the other ($\thicksim$200, 50) K.

\acknowledgements
We are grateful to the anonymous referee for his/ her comments that improved the paper.
This work is based on observations with \akari, a JAXA project with the participation
of ESA and makes use of data products from \hip\ Catalogs ( the primary result of the
Hipparcos space astrometry mission, undertaken by the European Space Agency), \mass\
(a joint project of the University of Massachusetts and the Infrared Processing and
Analysis Center /California Institute of Technology), \wise\ (a joint project of the
University of California, Los Angeles, and the Jet Propulsion Laboratory/California
Institute of Technology). This work makes use of the NASA/IPAC Infrared Science Archive,
which is operated by the Jet Propulsion Laboratory, California Institute of Technology,
under contract with the National Aeronautics and Space Administration.
This research makes use of ATLAS9 model and the SIMBAD database, operated at the
CDS, Strasbourg, France.

\clearpage

\clearpage

\clearpage

\clearpage

\clearpage
\begin{figure}
\includegraphics[width=6.5 in]{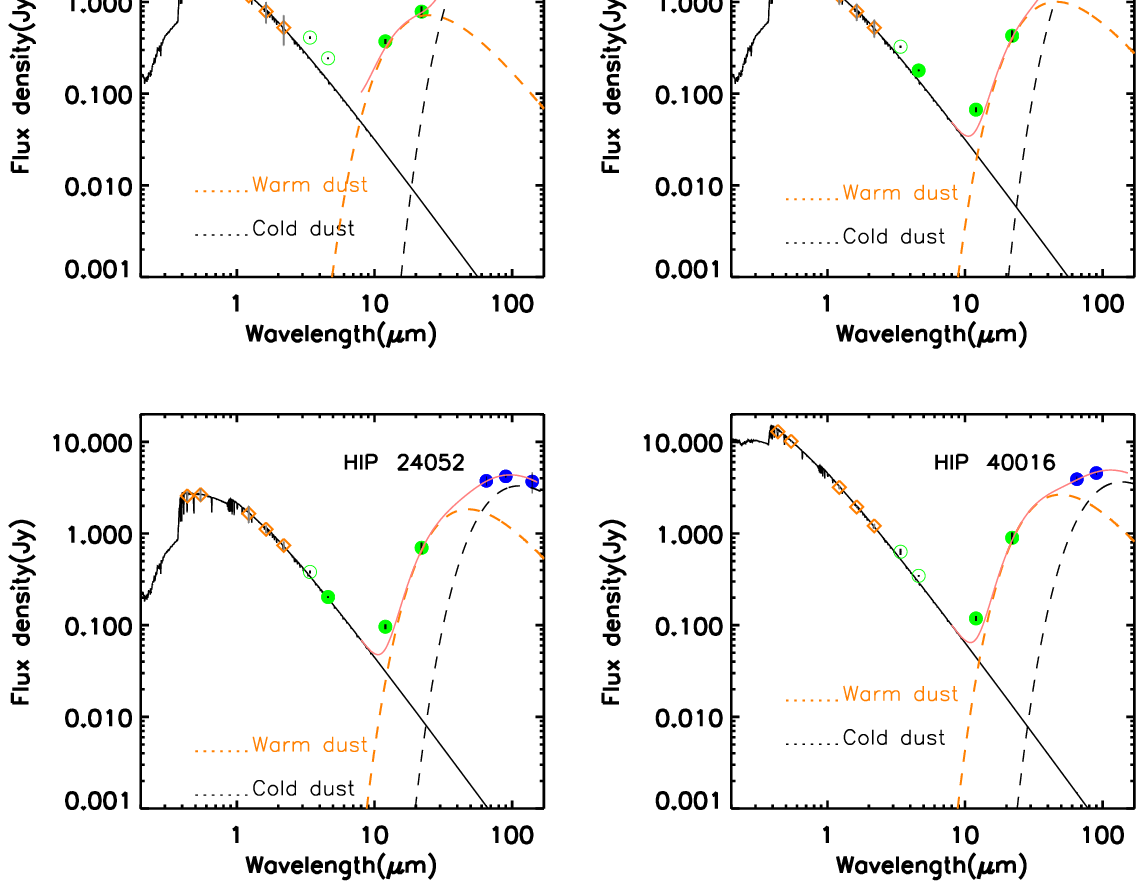}
\caption{SED fittings of the 4 stars that require double blackbody components. The symbols
legend as Figure \ref{bbmatch}. The orange dotted line is the fitted blackbody emission of
the warm dust.}
\label{bbmatch2}
\end{figure}

\clearpage
\begin{figure}
\plotone{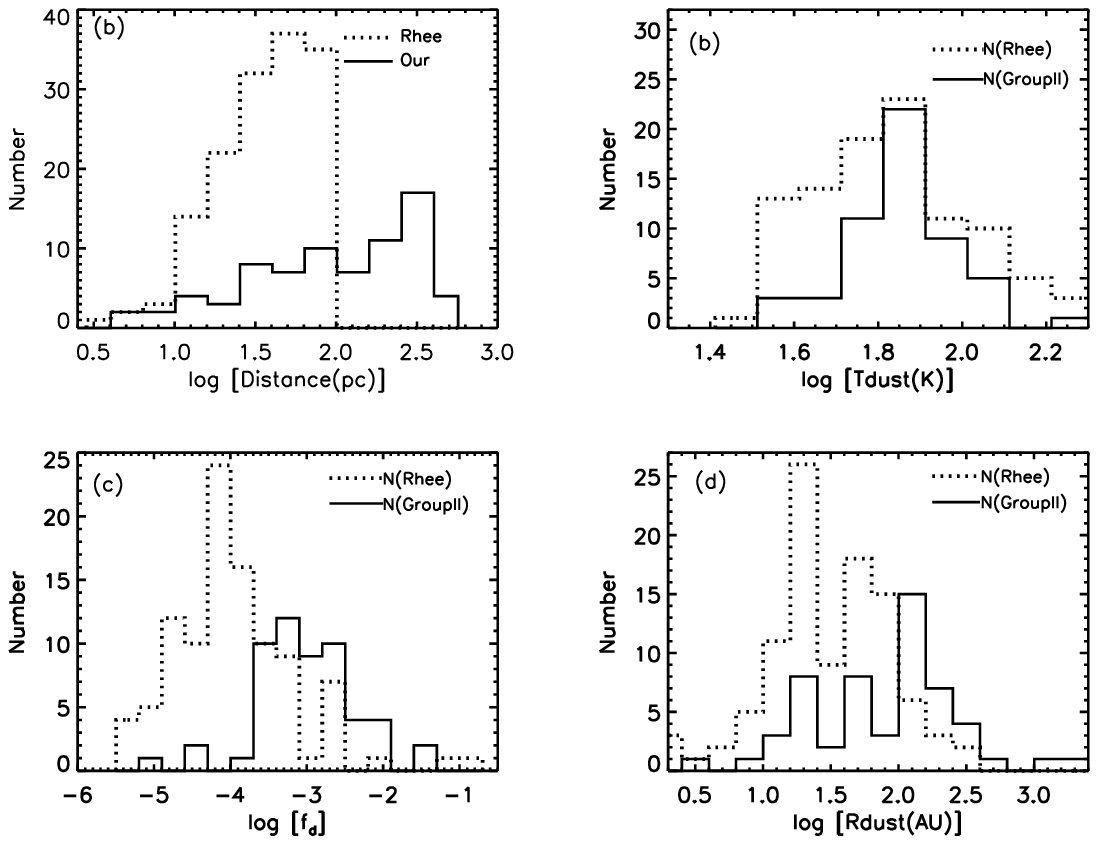}
\caption{The distributions of disk parameters for IR excess stars. Only Group II objects
were plotted in the panel b, c, d. The black solid line is our sample. The added dotted
line is the \iras\ sample \citep{rhe07}.}
\label{dusthist}
\end{figure}

\clearpage
\begin{figure}
\plotone{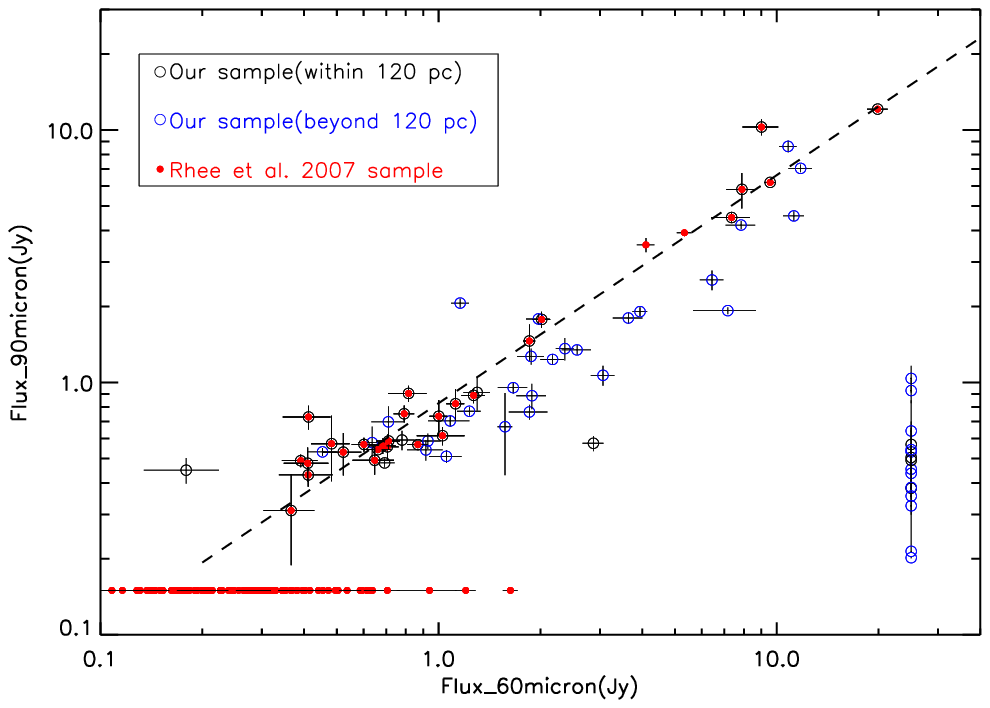}
\caption{A comparison of flux densities at 90 \micron\ from \fis\ and 60 \micron\ flux
from \iras. The hollow circles are our sample stars: black-- within 120 parsecs; blue--
beyond 120 parsecs. The red dots are Rhee's sample. Our sample stars without \iras\
detections are plotted on the right hand edge and Rhee's sample stars without \fis\ detections
are plotted on the downward edge. }
\label{f90}
\end{figure}


\clearpage
\begin{deluxetable}{ccc}
\tablenum{1}
\tablecolumns{3}
\tabletypesize{\scriptsize}
\tablecaption{The list of rejected sources \label{tb-reject}}
\tablewidth{0pt}
\tablehead{
 \colhead{HIP}           &\colhead{\fis\ identification} &\colhead{Reason for Rejection}
}
\startdata
        3401  &    0043182+615442  & 1   \\
        4023\tablenotemark{a}  &  0051337+513424    & 4   \\
        5147  &    0105535+655820  & 2   \\
       13330\tablenotemark{a}  &  0251319+674845    & 4   \\
       15984\tablenotemark{a}  &  0325506+305559    & 4   \\
       16826  &    0336292+481134  & 2   \\
       17890  &    0349363+385902  & 3   \\
       19395\tablenotemark{a}  &  0409164+304638    & 4   \\
       19720\tablenotemark{a}  &  0413352+101240    & 4   \\
       19762\tablenotemark{b}  &  0414129+281229    & 4   \\
       22910  &    0455460+303320  & 3   \\
       22925  &    0455593+303403  & 3   \\
       23143  &    0458465+295039  & 3   \\
       23428  &    0502065-712018  & 1   \\
       23633  &    0504502+264318  & 3   \\
       23734  &    0506086+585829  & 2   \\
       23873  &    0507494+302410  & 3   \\
       24552  &    0516006-094831  & 3   \\
       25253  &    0524009+245746  & 3   \\
       25258  &    0524079+022751  & 3   \\
       25299  &    0524426+014349  & 3   \\
       25793  &    0530272+251957  & 3   \\
       26295  &    0535587+244500  & 3   \\
       26451  &    0537385+210834  & 2   \\
       28582  &    0601597+163102  & 2   \\
       30089  &    0619582-103822  & 5   \\
       30800  &    0628177-130310  & 2   \\
       32349  &    0645085-164258  & 8   \\
       32677\tablenotemark{a}  & 0648585-150849  & 4   \\
       32923  &    0651333-065751  & 2   \\
       36369  &    0729106+205450  & 1   \\
       37279  &    0739178+051322  & 8   \\
       53691  &    1059071-770138  & 3   \\
       53911  &    1101516-344214  & 3   \\
       54413  &    1108017-773912  & 3   \\
       56379  &    1133251-701146  & 2   \\
       58520  &    1200066-781135  & 2   \\
       60936  &    1229071+020309  & 6   \\
       63973\tablenotemark{b}  &    1306360-494107  & 4   \\
       71352  &    1435303-420930  & 1   \\
       72685\tablenotemark{b}  &    1451400-305312  & 4   \\
       77542  &    1549578-035515  & 3   \\
       77952\tablenotemark{b}  &    1555094-632558  & 4   \\
       78034  &    1556019-660907  & 2   \\
       78317  &    1559283-402150  & 3   \\
       78594\tablenotemark{a}  &    1602491-044922  & 4   \\
       78943\tablenotemark{b}  &    1606579-274308  & 4   \\
       79080  &    1608344-390612  & 3   \\
       79476  &    1613116-222904  & 3   \\
       81624  &    1640176-235344  & 3   \\
       82747  &    1654450-365317  & 3   \\
       85792  &    1731503-495235  & 2   \\
       93975\tablenotemark{b}  &    1908039+214151  & 4   \\
       94260  &    1911115+154717  & 3   \\
       97649  &    1950472+085209  & 8   \\
      101983  &    2040025-603307  & 8   \\
      104580  &    2111024-634106  & 7   \\
      105638  &    2123489-404203  & 1   \\
      106079  &    2129147+442027  & 2   \\
      111785  &    2238316+555006  & 2   \\
      112377\tablenotemark{b}  &    2245378+415308  & 4   \\
\enddata
\tablecomments{
a: Stars in nebula.\\
b: Rejected by diffuse WISE images.\\
Col.(1): \hip\ identification. Col.(2): \fis\ identification.
Col.(3): Reason for rejection.
1. O star.
2. Be star.
3. Young stellar objects (YSOs) or PMS stars.
4. Contamination.
5. Post AGB star.
6. Quasar.
7. \fis\ flux density is not reliable (Fqual=1).
8. No FIR excess.
}
\end{deluxetable}

\clearpage
\begin{deluxetable}{ccccccccccccccccccccc}
\tablenum{2}
\tablecolumns{21}
\tabletypesize{\scriptsize}
\rotate
\tablecaption{The photometry and flux density for all sources \label{tb-flux}}
\tablewidth{0pt}
\tablehead{
 \colhead{Name}  &\multicolumn{2}{c}{Hipparcos} &\multicolumn{4}{c}{2MASS} &\multicolumn{8}{c}{WISE}
 &\multicolumn{6}{c}{AKARI/FIS} \\
\cmidrule(lr){2-3}\cmidrule(lr){4-6}\cmidrule(lr){7-15}
\cmidrule(lr){16-21} \colhead{HIP}  &\colhead{B}   &\colhead{V}   &\colhead{J}   &\colhead{H}
 &\colhead{K} &\colhead{rdflg} &\colhead{w1} &\colhead{w2} &\colhead{w3} &\colhead{w4}
 &\colhead{w1sat} &\colhead{w2sat} &\colhead{w3sat} &\colhead{w4sat} &\colhead{65\micron}
&\colhead{90\micron} &\colhead{140\micron} &\colhead{160\micron} &\colhead{FQUAL}
 &\colhead{offset}\\
   \colhead{}  &\colhead{mag}   &\colhead{mag}   &\colhead{mag}   &\colhead{mag}
 &\colhead{mag}  &\colhead{} &\colhead{mag} &\colhead{mag} &\colhead{mag} &\colhead{mag}
 &\colhead{\%} &\colhead{\%} &\colhead{\%} &\colhead{\%}
 &\colhead{Jy} &\colhead{Jy} &\colhead{Jy} &\colhead{Jy} &\colhead{} &\colhead{\arcsec}
}
\startdata
   746&   2.61&   2.27&   1.71&   1.58&   1.45&   333   &  -0.88 &  -0.18&   1.46 &   1.33 & 0.22  &  0.19  &  0.24  &  0.00   &   0.26 &   0.73 &    0.35 &    0.05 &  1311 &    3.1  \\
  4683&   8.65&   8.60&   7.87&   7.55&   7.45&   111   &   7.19 &   7.12&   4.73 &   2.56 & 0.07  &  0.00  &  0.00  &  0.00   &   7.17 &   8.61 &    7.78 &    6.26 &  3333 &    3.8  \\
  4789&   6.70&   6.70&   6.53&   6.56&   6.55&   111   &   6.52 &   6.47&   5.68 &   3.17 & 0.14  &  0.06  &  0.00  &  0.00   &   1.15 &   1.23 &    null &    0.57 &  1311 &   11.1  \\
  7345&   5.69&   5.62&   5.49&   5.53&   5.46&   111   &   5.47 &   5.30&   5.34 &   3.74 & 0.19  &  0.14  &  0.00  &  0.00   &   1.92 &   1.78 &    2.35 &    0.03 &  1311 &    2.7  \\
  7978&   6.08&   5.54&   4.79&   4.40&   4.34&   333   &   4.17 &   3.91&   4.22 &   3.95 & 0.21  &  0.22  &  0.00  &  0.00   &   1.44 &   0.90 &    0.56 &    null &  1311 &    7.0  \\
  8851&   9.58&   9.40&   8.99&   8.98&   8.95&   122   &   8.91 &   8.93&   8.35 &   6.11 & 0.01  &  0.00  &  0.00  &  0.00   &   0.32 &   0.53 &    1.50 &    null &  1311 &    8.4  \\
 10064&   3.14&   3.00&   2.74&   2.77&   2.68&   333   &   1.46 &   1.25&   2.68 &   2.46 & 0.20  &  0.19  &  0.09  &  0.00   &   0.42 &   0.59 &    null &    null &  1311 &    4.3  \\
 10670&   4.03&   4.01&   3.80&   3.86&   3.96&   331   &   3.95 &   3.64&   3.99 &   3.51 & 0.19  &  0.19  &  0.00  &  0.00   &   1.10 &   0.75 &    null &    0.49 &  1311 &    6.2  \\
 11847&   7.87&   7.47&   6.70&   6.61&   6.55&   111   &   6.54 &   6.52&   6.50 &   4.24 & 0.14  &  0.06  &  0.00  &  0.00   &   0.72 &   0.57 &    null &    null &  1311 &    5.2  \\
 13487&   8.84&   8.45&   7.66&   7.61&   7.57&   111   &   7.44 &   7.45&   6.59 &   3.22 & 0.06  &  0.00  &  0.00  &  0.00   &   4.95 &   7.05 &    3.08 &    3.60 &  3311 &    5.3  \\
 14043&   5.19&   5.24&   5.32&   5.40&   5.43&   111   &   5.33 &   5.26&   5.36 &   4.01 & 0.22  &  0.15  &  0.00  &  0.00   &   0.62 &   1.04 &    1.90 &    4.71 &  1311 &   14.8  \\
 16188&   7.39&   7.30&   6.56&   6.55&   6.49&   111   &   6.47 &   6.41&   6.32 &   5.27 & 0.13  &  0.06  &  0.00  &  0.00   &   1.54 &   0.76 &    0.89 &    1.39 &  1311 &   18.3  \\
 17812&   8.56&   8.45&   8.07&   8.08&   8.00&   111   &   8.02 &   8.00&   7.56 &   5.50 & 0.00  &  0.00  &  0.00  &  0.00   &   0.45 &   0.45 &    0.02 &    1.45 &  1311 &    8.3  \\
 17941&   8.93&   8.81&   8.60&   8.62&   8.58&   122   &   8.58 &   8.58&   7.50 &   4.24 & 0.00  &  0.00  &  0.00  &  0.00   &   1.02 &   1.36 &    2.04 &    1.54 &  1311 &   12.2  \\
 19475&   9.57&   9.30&   8.06&   8.01&   7.94&   111   &   7.88 &   7.90&   7.93 &   7.93 & 0.03  &  0.00  &  0.00  &  0.00   &   null &   0.20 &    null &    4.02 &  1113 &    9.4  \\
 20556&   7.02&   6.84&   6.31&   6.24&   6.26&   111   &   6.26 &   6.15&   5.97 &   4.78 & 0.14  &  0.09  &  0.00  &  0.00   &   0.25 &   0.77 &    0.55 &    0.12 &  1311 &    3.6  \\
 20884&   5.44&   5.54&   5.73&   5.79&   5.79&   111   &   5.87 &   5.75&   5.50 &   3.07 & 0.18  &  0.12  &  0.00  &  0.00   &   null &   0.57 &    null &    0.57 &  1311 &   18.2  \\
 21219&   7.06&   6.90&   6.52&   6.52&   6.48&   111   &   6.45 &   6.44&   6.50 &   6.43 & 0.13  &  0.06  &  0.00  &  0.00   &   null &   0.21 &    1.93 &    null &  1131 &    5.7  \\
 21898&   8.52&   8.20&   8.02&   7.99&   7.89&   111   &   7.82 &   7.82&   7.35 &   5.36 & 0.05  &  0.00  &  0.00  &  0.00   &   0.97 &   0.95 &    1.11 &    1.49 &  1311 &   16.9  \\
 22845&   4.73&   4.64&   4.85&   4.52&   4.42&   311   &   4.41 &   4.17&   4.43 &   4.06 & 0.23  &  0.23  &  0.00  &  0.00   &   0.23 &   0.43 &    0.64 &    3.41 &  1311 &    8.1  \\
 23451&   8.61&   8.50&   7.69&   7.62&   7.59&   111   &   7.59 &   7.63&   6.92 &   3.95 & 0.12  &  0.00  &  0.00  &  0.00   &   null &   0.82 &    null &    null &  1311 &    4.0  \\
 24052&   8.23&   8.10&   7.28&   7.35&   7.30&   111   &   7.27 &   7.31&   6.20 &   2.69 & 0.08  &  0.00  &  0.00  &  0.00   &   3.75 &   4.20 &    3.70 &    0.93 &  3331 &    9.9  \\
 26062&   7.00&   6.97&   6.84&   6.92&   6.82&   111   &   6.81 &   6.75&   5.13 &   2.29 & 0.11  &  0.04  &  0.00  &  0.00   &   1.44 &   0.91 &    null &    1.29 &  1311 &    4.1  \\
 27296&   7.14&   7.12&   7.09&   7.13&   7.12&   111   &   7.10 &   7.14&   6.71 &   3.91 & 0.09  &  0.00  &  0.00  &  0.00   &   0.87 &   1.35 &    1.34 &    0.83 &  1311 &    5.0  \\
 27321&   4.02&   3.85&   3.67&   3.54&   3.53&   333   &   3.48 &   3.18&   2.60 &   0.01 & 0.24  &  0.24  &  0.09  &  0.00   &  15.72 &  12.10 &    5.88 &    2.95 &  3331 &    4.2  \\
 32345&   7.44&   7.45&   7.50&   7.53&   7.52&   111   &   7.47 &   7.51&   7.30 &   5.96 & 0.06  &  0.00  &  0.00  &  0.00   &   0.64 &   0.59 &    2.18 &    0.66 &  1311 &   10.5  \\
 36437&   7.10&   7.18&   7.30&   7.43&   7.38&   111   &   7.33 &   7.43&   7.02 &   3.68 & 0.08  &  0.00  &  0.00  &  0.00   &   0.70 &   0.64 &    0.00 &    null &  1311 &   11.2  \\
 36581&   8.12&   7.95&   7.82&   7.77&   7.69&   111   &   7.76 &   7.60&   6.49 &   4.70 & 0.05  &  0.00  &  0.00  &  0.00   &   0.94 &   0.53 &    null &    1.55 &  1311 &    7.7  \\
 40016&   6.32&   6.47&   6.72&   6.84&   6.83&   111   &   6.71 &   6.74&   5.97 &   2.41 & 0.13  &  0.03  &  0.00  &  0.00   &   3.91 &   4.57 &    2.15 &    0.68 &  3311 &    3.5  \\
 40024&   7.85&   7.93&   8.04&   8.06&   8.06&   111   &   8.05 &   8.05&   7.37 &   4.05 & 0.00  &  0.00  &  0.00  &  0.00   &   1.30 &   1.78 &    1.22 &    1.24 &  1311 &    4.1  \\
 40748&  10.38&  10.40&  10.32&  10.31&  10.25&   222   &  10.15 &   9.95&   8.57 &   5.72 & 0.00  &  0.00  &  0.00  &  0.00   &   0.22 &   0.44 &    null &    null &  1311 &   13.2  \\
 41650&   8.60&   8.60&   8.43&   8.28&   7.92&   111   &   7.04 &   6.30&   3.09 &   1.13 & 0.24  &  0.15  &  0.08  &  0.00   &   1.41 &   1.27 &    1.48 &    0.23 &  1311 &    5.9  \\
 44001&   5.87&   5.66&   5.27&   5.21&   5.16&   111   &   5.16 &   4.98&   5.20 &   4.89 & 0.21  &  0.18  &  0.00  &  0.00   &   0.17 &   0.49 &    null &    1.48 &  1311 &    6.3  \\
 45581&   5.30&   5.28&   5.24&   5.27&   5.17&   111   &   5.06 &   4.90&   5.14 &   4.84 & 0.20  &  0.14  &  0.00  &  0.00   &   0.39 &   0.67 &    0.50 &    null &  1311 &   11.8  \\
 46021&   8.98&   8.90&   8.62&   8.66&   8.59&   112   &   8.53 &   8.53&   7.59 &   5.50 & 0.00  &  0.00  &  0.00  &  0.00   &   null &   0.58 &    1.93 &    1.22 &  1311 &    8.6  \\
 48613&   5.71&   5.72&   5.70&   5.76&   5.74&   111   &   5.71 &   5.61&   5.66 &   4.56 & 0.18  &  0.14  &  0.00  &  0.00   &   0.99 &   0.48 &    null &    null &  1311 &   18.3  \\
 53524&   7.60&   7.36&   6.91&   6.87&   6.79&   111   &   6.72 &   6.75&   6.69 &   5.47 & 0.10  &  0.02  &  0.00  &  0.00   &   0.62 &   0.57 &    1.29 &    0.20 &  1311 &    2.7  \\
 55505&   9.72&   8.52&   6.40&   5.76&   5.59&   111   &   5.50 &   5.34&   3.11 &   0.20 & 0.20  &  0.15  &  0.07  &  0.00   &   7.18 &   5.82 &    2.80 &    2.57 &  3311 &    8.5  \\
 57632&   2.23&   2.14&   1.85&   1.93&   1.88&   333   &   0.46 &   0.13&   2.06 &   1.70 & 0.24  &  0.23  &  0.24  &  0.00   &   0.38 &   0.61 &    0.22 &    2.20 &  1311 &    7.3  \\
 57757&   4.15&   3.60&   2.60&   2.36&   2.27&   333   &   0.71 &   0.83&   2.39 &   2.29 & 0.24  &  0.22  &  0.21  &  0.00   &   0.15 &   0.45 &    null &    2.39 &  1311 &   15.8  \\
 60074&   7.63&   7.03&   5.87&   5.61&   5.54&   111   &   5.52 &   5.36&   5.54 &   5.18 & 0.18  &  0.13  &  0.00  &  0.00   &   0.62 &   0.56 &    null &    null &  1311 &    9.4  \\
 61498&   5.78&   5.78&   5.78&   5.79&   5.77&   111   &   5.37 &   5.40&   5.02 &   1.22 & 0.17  &  0.10  &  0.00  &  0.00   &   6.07 &   4.50 &    3.26 &    null &  3311 &    8.9  \\
 65875&   8.58&   8.08&   7.17&   6.97&   6.90&   111   &   6.86 &   6.86&   6.70 &   3.99 & 0.20  &  0.00  &  0.00  &  0.00   &   0.31 &   0.57 &    null &    null &  1311 &    6.7  \\
 73145&   8.05&   7.90&   7.60&   7.56&   7.52&   111   &   7.51 &   7.52&   6.92 &   4.27 & 0.07  &  0.00  &  0.00  &  0.00   &   0.60 &   0.56 &    0.07 &    null &  1311 &    5.4  \\
 74421&   6.02&   6.01&   5.91&   5.91&   5.91&   111   &   5.91 &   5.81&   5.82 &   5.26 & 0.17  &  0.11  &  0.00  &  0.01   &   0.81 &   2.06 &    4.51 &    5.28 &  1331 &   16.4  \\
 76736&   6.51&   6.43&   6.30&   6.34&   6.27&   111   &   6.27 &   6.22&   6.16 &   5.01 & 0.15  &  0.08  &  0.00  &  0.00   &   0.45 &   0.57 &    null &    0.17 &  1311 &   10.4  \\
 76829&   5.04&   4.62&   4.02&   3.73&   3.80&   333   &   3.68 &   3.09&   3.65 &   3.52 & 0.25  &  0.23  &  0.02  &  0.00   &   1.23 &   0.54 &    null &    1.71 &  1311 &    9.0  \\
 77441&   8.57&   8.10&   7.39&   7.22&   7.20&   111   &   7.16 &   7.18&   7.07 &   6.35 & 0.11  &  0.00  &  0.00  &  0.00   &   0.30 &   0.38 &    null &    0.36 &  1311 &    5.7  \\
 79977&   9.58&   9.09&   8.06&   7.85&   7.80&   111   &   7.76 &   7.76&   7.46 &   4.29 & 0.08  &  0.00  &  0.00  &  0.00   &   0.63 &   0.70 &    null &    1.04 &  1311 &    6.9  \\
 80951&  10.04&   9.40&   8.52&   8.32&   8.26&   112   &   8.20 &   8.23&   8.19 &   7.93 & 0.00  &  0.00  &  0.00  &  0.00   &   0.12 &   0.38 &    1.40 &    null &  1311 &    6.9  \\
 81474&   6.84&   6.70&   5.90&   5.78&   5.69&   111   &   5.61 &   5.50&   5.45 &   3.79 & 0.20  &  0.12  &  0.00  &  0.00   &   1.70 &   1.92 &    null &    0.06 &  1311 &   12.5  \\
 81891&   6.38&   6.46&   6.58&   6.67&   6.63&   111   &   6.55 &   6.62&   6.29 &   4.45 & 0.12  &  0.04  &  0.00  &  0.00   &   0.52 &   0.70 &    null &    0.12 &  1311 &    8.5  \\
 82770&   8.46&   7.95&   6.96&   6.75&   6.64&   111   &   6.61 &   6.59&   6.67 &   6.54 & 0.14  &  0.04  &  0.00  &  0.00   &   0.58 &   0.35 &    0.60 &    null &  1311 &   15.7  \\
 83505&   8.24&   8.10&   7.52&   7.54&   7.52&   111   &   7.47 &   7.51&   7.24 &   5.23 & 0.15  &  0.00  &  0.00  &  0.00   &   1.25 &   0.89 &    null &    2.87 &  1311 &    4.0  \\
 85537&   5.62&   5.39&   4.81&   4.88&   4.80&   111   &   4.78 &   4.57&   4.80 &   4.60 & 0.22  &  0.22  &  0.00  &  0.00   &   0.16 &   0.31 &    null &    null &  1311 &   10.1  \\
 86078&   8.00&   7.80&   7.02&   6.95&   6.79&   111   &   6.71 &   6.70&   6.72 &   6.08 & 0.12  &  0.03  &  0.00  &  0.00   &   0.47 &   0.54 &    1.91 &    null &  1311 &   19.1  \\
 87108&   3.79&   3.75&   3.59&   3.66&   3.62&   333   &   3.68 &   3.36&   3.65 &   3.12 & 0.26  &  0.24  &  0.03  &  0.00   &   1.31 &   0.89 &    null &    0.59 &  1301 &    5.7  \\
 87807&   7.94&   7.70&   7.45&   7.45&   7.39&   111   &   7.34 &   7.37&   6.97 &   5.49 & 0.18  &  0.00  &  0.00  &  0.00   &   0.08 &   0.51 &    null &    0.42 &  1311 &    6.0  \\
 88399&   7.46&   7.01&   6.16&   6.02&   5.91&   111   &   5.76 &   5.70&   5.80 &   4.94 & 0.18  &  0.10  &  0.00  &  0.00   &   0.60 &   0.49 &    0.45 &    null &  1311 &    4.7  \\
 88983&   8.15&   8.00&   7.30&   7.19&   7.18&   111   &   7.16 &   7.17&   7.22 &   7.07 & 0.09  &  0.00  &  0.00  &  0.00   &   0.26 &   0.50 &    2.25 &    null &  1311 &   18.3  \\
 90491&   8.76&   8.50&   8.23&   8.21&   8.13&   111   &   8.09 &   8.10&   7.97 &   6.12 & 0.00  &  0.00  &  0.00  &  0.00   &   null &   0.49 &    1.03 &    0.51 &  1311 &    4.5  \\
 91262&   0.03&   0.03&  -0.18&  -0.03&   0.13&   333   &  -2.03 &  -2.08&   0.02 &  -0.16 & 0.31  &  0.31  &  0.30  &  0.00   &   6.58 &   6.20 &    4.05 &    3.22 &  3311 &    3.4  \\
 92800&   6.80&   6.80&   6.53&   6.54&   6.50&   111   &   6.42 &   6.42&   6.17 &   4.05 & 0.13  &  0.04  &  0.00  &  0.00   &   2.71 &   2.55 &    2.19 &    3.02 &  3311 &    6.4  \\
 93000&   7.37&   7.15&   6.63&   6.61&   6.58&   111   &   6.54 &   6.46&   6.05 &   4.16 & 0.13  &  0.05  &  0.00  &  0.00   &   1.17 &   1.80 &    0.78 &    null &  1311 &   11.9  \\
 95270&   7.52&   7.04&   6.20&   5.98&   5.91&   111   &   5.89 &   5.81&   5.89 &   3.95 & 0.18  &  0.11  &  0.00  &  0.00   &   1.71 &   1.46 &    1.24 &    1.23 &  1311 &    7.9  \\
 95619&   5.64&   5.66&   5.67&   5.66&   5.68&   111   &   5.68 &   5.56&   5.61 &   4.58 & 0.21  &  0.13  &  0.00  &  0.00   &   0.45 &   0.73 &    0.07 &    null &  1311 &    7.6  \\
 99273&   7.66&   7.18&   6.32&   6.09&   6.08&   111   &   6.06 &   5.88&   6.00 &   4.06 & 0.16  &  0.07  &  0.00  &  0.00   &   1.22 &   0.59 &    1.48 &    0.15 &  1311 &   14.6  \\
101612&   5.04&   4.75&   4.28&   4.02&   4.04&   333   &   4.06 &   3.58&   4.04 &   3.87 & 0.24  &  0.22  &  0.00  &  0.00   &   null &   0.53 &    null &    0.37 &  1311 &    4.9  \\
102761&   8.01&   7.97&   7.89&   7.93&   7.95&   111   &   7.92 &   7.95&   7.13 &   4.06 & 0.03  &  0.00  &  0.00  &  0.00   &   1.66 &   1.91 &    1.59 &    0.01 &  1311 &   11.0  \\
104354&   8.37&   8.31&   8.12&   8.13&   8.12&   112   &   8.10 &   8.13&   7.21 &   4.70 & 0.00  &  0.00  &  0.00  &  0.00   &   0.10 &   0.54 &    null &    null &  1311 &    7.3  \\
111214&   9.16&   8.90&   8.58&   8.64&   8.62&   122   &   8.59 &   8.62&   8.19 &   6.50 & 0.00  &  0.00  &  0.00  &  0.00   &   0.20 &   0.32 &    1.12 &    0.30 &  1311 &   12.8  \\
111429&   6.88&   7.00&   7.19&   7.27&   7.29&   111   &   7.23 &   7.34&   7.38 &   6.68 & 0.08  &  0.00  &  0.00  &  0.00   &   0.10 &   0.93 &    2.63 &    1.26 &  1311 &   15.2  \\
113368&   1.25&   1.17&   1.04&   0.94&   0.94&   333   &  -1.47 &  -0.75&   1.11 &   0.79 & 0.31  &  0.28  &  0.29  &  0.00   &   8.30 &  10.27 &    7.71 &    4.66 &  3131 &    4.9  \\
114189&   6.25&   5.98&   5.38&   5.28&   5.24&   111   &   5.19 &   5.04&   5.21 &   4.85 & 0.20  &  0.16  &  0.00  &  0.00   &   1.12 &   0.48 &    0.26 &    3.48 &  1311 &    4.8  \\
118289&   7.13&   7.17&   7.18&   7.26&   7.26&   111   &   7.18 &   7.26&   6.22 &   3.64 & 0.06  &  0.00  &  0.00  &  0.00   &   1.13 &   1.07 &    1.09 &    null &  1311 &    6.3  \\
\enddata
\tablecomments{
Col.(1): \hip\ identification. Col.(2): B magnitude. Col.(3): V magnitude. Col.(4): J magnitude.
Col.(5): H magnitude.Col.(6): K magnitude.  Col.(7): Col. Read flag.(8): \wise\ W1 magnitude. Col.(9): \wise\ W2 magnitude. Col.(10): \wise\ W3 magnitude. Col.(11): \wise\ W4 magnitude. Col.(12): Saturated pixel fraction, W1.
Col.(13): Saturated pixel fraction, W2. Col.(14): Saturated pixel fraction, W3.
Col.(15): Saturated pixel fraction, W4. Col.(16): \fis\ 65 \micron\ flux density.
Col.(17): \fis\ 90 \micron\ flux density. Col.(18): \fis\ 140 \micron\ flux density.
Col.(19): \fis\ 160 \micron\ flux density.
Col.(20):Flux density quality flag in \fis\ 4 bands:
3=High quality (the source is confirmed and flux is reliable);
2=The source is confirmed but flux is not reliable (see FLAGS);
1=The source is not confirmed;
0=Not observed (no scan data available).
Col.(21): \fis\ position offset.
}
\end{deluxetable}

\clearpage
\begin{deluxetable}{ccccccccccc}
\tablenum{3}
\tablecolumns{11}
\tabletypesize{\scriptsize}
\rotate
\tablecaption{The star basic properties and dust basic properties of our sample sources (Group II)\label{tb-dust}}
\tablewidth{0pt}
\tablehead
{
\colhead{HIP}   &\colhead{Distance(pc)}  &\colhead{$T_{\rm eff}$}
&\colhead{log{\it g}}   &\colhead{E(B-V)}  &\colhead{$T_{\rm d}$ (K)}
&\colhead{$R_{\rm d}$ (AU)}  &\colhead{$M_{\rm d}$($M_\earth$)}   & \colhead{$f_{\rm d}$}
&\colhead{Sp.Type}    &\colhead{References} \\
\colhead{(1)}        &\colhead{(2)}        &\colhead{(3)}        &
\colhead{(4)}        &\colhead{(5)}        &\colhead{(6)}        &
\colhead{(7)}        &\colhead{(8)}        &\colhead{(9)}        &
\colhead{(10)}       &\colhead{(11)}
}
\startdata
     4683\tablenotemark{\dag} &  446.4  &   10000 &   4.0 & 0.200 &    53 &  363 & 2.22e+02 & 1.53e-02  &       B5 & 1                      \\
     4789 &  357.1  &   11000 &   4.0 & 0.093 &    106 &   145 & 3.59e+00 & 1.55e-03 &       B9 & 1,4                    \\
     7345 &   61.3  &    9500 &   4.5 & 0.028 &     77 &    60 & 3.05e-01 & 7.65e-04 &      A1V & 1,2,4,10               \\
     8851 &  411.5  &   10500 &   4.0 & 0.200 &    136 &    31 & 5.47e-02 & 4.89e-04 &       B8 & ...                    \\
    11847 &   63.7  &    8000 &   4.5 & 0.178 &     90 &    22 & 7.46e-02 & 1.36e-03 &       F0 & 2,5,10,11             \\
    13487\tablenotemark{\dag} &  261.1  &   10000 &   4.0 & 0.200 &    41 &  367 & 1.55e+02 & 1.05e-02  &       B8 & 1                      \\
    14043 &  243.9  &   20000 &   4.0 & 0.138 &     81 &   701 & 2.54e+00 & 4.71e-05 &      B7V & ...                    \\
    16188 &  353.4  &    8750 &   4.0 & 0.192 &     69 &   250 & 5.80e+00 & 8.43e-04 &       A0 & ...                    \\
    17812 &  393.7  &   10000 &   4.0 & 0.157 &     79 &   128 & 2.92e+00 & 1.61e-03 &       B9 & ...                    \\
    17941 &  331.1  &    9500 &   4.5 & 0.064 &     86 &    63 & 5.06e+00 & 1.14e-02 &       A0 & ...                    \\
    20556 &  194.9  &    9750 &   4.0 & 0.200 &     75 &   154 & 1.43e+00 & 5.48e-04 &       A2 & ...                    \\
    20884 &  118.1  &   13000 &   4.0 & 0.000 &    119 &    67 & 1.46e-01 & 2.96e-04 &      B3V & 1                    \\
    21898 &  537.6  &    8250 &   4.5 & 0.000 &     70 &   169 & 1.63e+01 & 5.17e-03 &       A0 & ...                    \\
    23451 &  112.1  &    8500 &   4.0 & 0.200 &     96 &    23 & 2.75e-01 & 4.56e-03 &       A0 & 2,10                   \\
    24052\tablenotemark{\dag} &  159.2  &   10000 &   4.0 & 0.200 &    45 &  214 & 1.76e+01 & 3.48e-03  &       B9 & ...                    \\
    26062 &  100.0  &   10000 &   4.0 & 0.048 &    134 &    18 & 1.33e-01 & 3.40e-03 &       B8 & ...                    \\
    27296 &  325.7  &   15000 &   4.0 & 0.134 &     85 &   247 & 5.18e+00 & 7.69e-04 &       B8 & ...                    \\
    27321 &   19.3  &    8500 &   4.5 & 0.010 &    120 &    16 & 8.08e-02 & 2.85e-03 &      A3V & 1,2,6,7,10,12,13     \\
    32345 &  454.5  &   10500 &   4.0 & 0.002 &     67 &   272 & 8.01e+00 & 9.86e-04 &       B9 & ...                    \\
    36437 &  490.2  &   14000 &   4.0 & 0.034 &     98 &   219 & 4.13e+00 & 7.84e-04 &   B3IV/V & ...                    \\
    36581 &  666.7  &    8750 &   4.5 & 0.011 &     93 &   142 & 7.12e+00 & 3.22e-03 &       F8 & ...                    \\
    40016\tablenotemark{\dag} &  500.0  &   16000 &   4.0 & 0.008 &    37 & 2500 & 3.48e+02 & 5.07e-04  &     B3IV & 1                      \\
    40024 &  450.5  &   11500 &   4.0 & 0.003 &     82 &   158 & 1.43e+01 & 5.18e-03 &      B6V & ...                    \\
    40748 &  763.4  &   10000 &   4.0 & 0.036 &     86 &    71 & 7.56e+00 & 1.35e-02 &       B4 & ...                    \\
    41650 &  288.2  &    9250 &   4.0 & 0.101 &    194 &    12 & 8.48e-01 & 4.73e-02 &     A0IV & 1                      \\
    46021 &  458.7  &    9750 &   4.0 & 0.098 &     77 &   112 & 5.23e+00 & 3.78e-03 & A0III/IV & ...                    \\
    48613 &   97.8  &   10000 &   4.0 & 0.000 &     83 &    80 & 1.79e-01 & 2.50e-04 &      A0V & 4                      \\
    53524 &   91.6  &    9250 &   4.5 & 0.169 &     70 &    58 & 2.79e-01 & 7.41e-04 &    A8III & 2,4                    \\
    55505 &   46.7  &    4750 &   4.0 & 0.184 &    139 &     4 & 1.67e-01 & 6.16e-02 &      K4V & 2                      \\
    61498 &   67.1  &   10000 &   4.5 & 0.001 &    101 &    36 & 5.26e-01 & 3.68e-03 &      A0V & 2,3,10,13             \\
    65875 &   97.1  &    7250 &   4.5 & 0.154 &     95 &    21 & 1.52e-01 & 2.90e-03 &      F6V & ...                    \\
    73145 &  111.1  &    9250 &   4.5 & 0.096 &     97 &    25 & 1.86e-01 & 2.53e-03 &     A2IV & 2,10                   \\
    74421 &  260.4  &   10250 &   4.0 & 0.053 &     27 &  1851 & 2.25e+02 & 5.97e-04 & B8/B9III & ...                    \\
    76736 &   77.3  &    9500 &   4.5 & 0.030 &     75 &    55 & 1.68e-01 & 4.95e-04 &      A5V & 2                      \\
    77441 &  117.0  &    7500 &   4.5 & 0.109 &     63 &    55 & 4.03e-01 & 1.18e-03 &   F2/F3V & ...                    \\
    79977 &  131.8  &    7250 &   4.5 & 0.190 &     87 &    23 & 4.17e-01 & 6.77e-03 &   F2/F3V & 4                      \\
    81474 &  165.0  &   10000 &   4.0 & 0.200 &     78 &   144 & 2.32e+00 & 1.01e-03 &   B9.5IV & 1                      \\
    81891 &  240.4  &   12000 &   4.0 & 0.001 &     87 &   155 & 1.43e+00 & 5.37e-04 &      B8V & ...                    \\
    83505 &  342.5  &   10000 &   4.0 & 0.200 &     71 &   175 & 5.86e+00 & 1.74e-03 &  B9.5III & ...                    \\
    86078 &  317.5  &    8250 &   4.0 & 0.197 &     62 &   215 & 4.43e+00 & 8.70e-04 &       A0 & ...                    \\
    87807 &  207.9  &   11750 &   4.0 & 0.200 &     74 &   129 & 1.11e+00 & 6.04e-04 &       B9 & ...                    \\
    88399 &   46.9  &    7750 &   4.5 & 0.194 &     76 &    29 & 5.16e-02 & 5.54e-04 &      F5V & 2,5,10,12              \\
    90491 &  117.5  &    8250 &   4.5 & 0.008 &     67 &    35 & 4.41e-01 & 3.13e-03 &       A0 & ...                    \\
    91262 &    7.8  &   10000 &   4.0 & 0.022 &     65 &   146 & 2.45e-02 & 1.04e-05 &      A0V & 1,2,6,7,13              \\
    92800 &  225.7  &   10000 &   4.0 & 0.095 &     72 &   174 & 7.04e+00 & 2.10e-03 &       A0 & 1                      \\
    93000 &  354.6  &   10000 &   4.0 & 0.200 &     75 &   243 & 1.09e+01 & 1.68e-03 & B8II/III & ...                    \\
    95270 &   50.6  &    6500 &   4.0 & 0.001 &     78 &    22 & 1.68e-01 & 3.05e-03 &   F5/F6V & 2,10,12              \\
    95619 &   69.1  &   10250 &   4.0 & 0.009 &     76 &    72 & 1.66e-01 & 2.90e-04 &   B8/B9V & 2,10                   \\
    99273 &   53.5  &    6500 &   4.0 & 0.001 &     91 &    16 & 5.22e-02 & 1.82e-03 &      F5V & 2,9,10               \\
   102761 &  431.0  &   18000 &   4.0 & 0.194 &     81 &   334 & 1.45e+01 & 1.18e-03 &       B8 & ...                   \\
   104354 &  555.6  &   14000 &   4.0 & 0.184 &     98 &   185 & 3.68e+00 & 9.74e-04 &       B9 & ...                  \\
   111214 &  393.7  &   11500 &   4.0 & 0.200 &     69 &   160 & 3.07e+00 & 1.09e-03 &       B9 & ...                  \\
   111429 &  289.9  &   14000 &   4.0 & 0.015 &     53 &   462 & 9.71e+00 & 4.14e-04 &    B1.5V & ...                  \\
   113368 &    7.7  &    8750 &   4.0 & 0.008 &     54 &   107 & 6.30e-02 & 4.98e-05 &      A3V & 1,2,5,6,7,8,13        \\
   118289 &  300.3  &   18000 &   4.0 & 0.142 &    103 &   195 & 2.24e+00 & 5.32e-04 &       B9 & 1                      \\

\enddata
\tablecomments{
\tablenotemark{\dag} SED fitting with 2 blackbody model.\\
Note. -- Col.(1): \hip\ identification.
Col.(2): Distance. Col.(3): Effective temperature. Col.(4): Surface gravity. Col.(5): E(B-V).
Col.(6): Dust temperature. Col.(7): Dust location. Col.(8): Total dust mass($M_\earth$).
Col.(9): Dust fractional luminosity. Col.(10): Spectral type. Col.(11): References --
(1)Oudmijer et al. 1992; (2)Rhee et al. 2007; (3)Koerner et al. 1998; (4)Chen et al. 2006;
(5)Decin et al. 2003; (6)Habing et al. 1999; (7)Habing et al. 2001; (8)Su et al. 2006;
(9)Carpenter et al. 2008; (10)Moo\'r et al. 2006; (11)Moo\'r et al. 2011; (12)Rebull et al. 2008
(13)Rieke et al. 2005 \\
}
\end{deluxetable}


\end{document}